\documentstyle[11pt,fleqn]{article}
\title{Comments on "Bigger Rip with no dark energy"} 
\author{A. Kwang-Hua CHU} 
\date{Department of Physics, Xinjiang University, 
Urumqi 830046, PR China}
\begin{document}           
\maketitle
\begin{abstract}
We make corrections on the paper by Frampton and Takahashi [{\it Astropart. Phys.}
{\bf 22} (2004) 307]. Our focus is  especially upon Eqs. (12-15) therein. \newline

\end{abstract}
\doublerulesep=6mm    %
\baselineskip=6mm
\bibliographystyle{plain}               

\noindent
Frampton and Takahashi just presented an interesting paper which shows that a
singularity can occur at a finite future time in an expanding
Friedmann universe [1].
By studying a modified Friedmann equation which arises in an extension of
general relativity (GR) or the modification of GR (due to Dvali,
Gabadadze and Porrati [2] (DGP); considering the
four-dimensional gravity which arises from five-dimensional
general relativity confined to a brane
with three space dimensions; the action is $S=[M(t)]^3 \int d^5 X \sqrt{G} {\cal R}^{(5)}$$+
M^2_{Planck} \int d^4 x \sqrt{g} R$, where ${\cal R}^{(5)}$ and $R$ are the scalar curvature in fiveand four-dimensional spacetime respectively, and
$G$ and $g$ are the determinant of the five- and
four-dimensional spacetime metric) which accommodates a
time-dependent fundamental length $L(t) (\equiv M^2_{Planck}/[M(t)]^3 = [H(t)]^{-1}$, they considered cosmological models where the scale factor diverges with an
essential singularity at a finite future time. Their models have no dark energy in the conventional sense of energy possessing
a simple pressure¨Cenergy relationship.  \newline
Frampton and Takahashi  generalized the Schwarzschild
solution to this modification of GR [7] together with the modification of
the Newton potential at
short distances being given by
\begin{displaymath}
 V(r)=-\frac{G m}{r}-\frac{4\sqrt{G m}\sqrt{r}}{L(t)}=-\frac{r_g}{2r}-\frac{2\sqrt{2}\sqrt{r_g r}}{L(t)},
\end{displaymath}
where $r_g = 2Gm$ is the Schwarzschild radius. \newline
The present author would like to point out some mistakes in [1] here.
Firstly, Frampton and Takahashi defined
$L(t)=[L(t_0)]^{-1} [T(t)]^p$ (cf. Eq. (6) in [1]) and then wrote down the modified Friedmann equation for DGP gravity being, in an empty universe, $[H(t)]^2-H(t)/L(t)=0$ (cf. Eq. (11) in [1]). He thus got
\begin{equation}
 \frac{\dot{a}(t)}{a(t)}=H(t)=\frac{H(t_0)}{T^p}
\end{equation}
which is Eq. (12) in [1]. But, as the present author checked, with $L(t)=1/[H(t)]$,
Eq. (12) should be read as
\begin{equation}
  \frac{\dot{a}(t)}{a(t)}=H(t)=\frac{1}{H(t_0) T^p}.
\end{equation}
Even if we follow their assumption : neglecting, for the future
evolution, the term $(\rho_M + \rho_{\gamma})/(3M^2_{Planck})$ on the
right-hand-side of the modified Friedmann equation. Meanwhile, with $\gamma= -dT/dt$$=
1/(t_{rip}-t_0)$, we then have
\begin{equation}
 \ln a(t)=-\int_1^{T(t)} \frac{L(t_0)}{\gamma T^p} dT,
\end{equation}
instead of the expression shown in the right-hand-side of Eq. (13) in [1]. Thus,
for $p = 1$, which is similar to dark energy
with a constant $w < - 1$ equation of state : Eq. (14) in [1] should be changed to
\begin{equation}
 a(t)=T^{-L(t_0)/\gamma},
\end{equation}
while for the Bigger Rip case $p > 1$ one finds
\begin{equation}
 a(t)=a(t_0) \exp[(\frac{1}{T^{p-1}}-1)\frac{L(t_0)}{\gamma(p-1)}],
\end{equation}
instead of the Eq. (15) shown in [1]. \newline
To conclude in brief, as the derivations of Eqs. (12-15) in [1] had something wrong
with the position of $L(t_0)$ (if Eq. (6) is correct in [1]) which the present author just corrected above, the remaining
statements and mathematical derivations associated with Eqs. (12-15) should be corrected
simultaneously for the physical interpretation of Frampton and Takahashi's
subsequent presentation in [1]. {\it Acknowledgements.} The author is partially supported
by the Starting Funds of 2005-XJU-Scholars.

\end{document}